\documentclass[preprint,showpacs,preprintnumbers]{revtex4}

\usepackage{amsmath}
\usepackage{graphicx}% Include figure files
\begin{document}

\title{Non-adiabatic optomechanical Hamiltonian of a moving dielectric
membrane in a cavity}

% Force line breaks with \\

\author{H. K. Cheung and C. K. Law}
\affiliation{Department of Physics and Institute of Theoretical
Physics, The Chinese University of Hong Kong, Shatin, Hong Kong SAR,
China}

\date{\today}
\begin{abstract}
We formulate a non-relativistic Hamiltonian in order to describe
the interaction between a moving dielectric membrane and radiation
pressure. Such a Hamiltonian is derived without making use of the
single-mode adiabatic approximation and linear approximation, and
hence it enables us to incorporate multi-mode effects and general
(non-relativistic) motion of the membrane in cavity optomechanics.
By performing second quantization, we show how a set of
generalized Fock states can be constructed to represent quantum
states of the membrane and cavity field. In addition, we discuss
examples showing how photon scattering among different cavity
modes would modify the interaction strengths and the mechanical
frequency of the membrane.
\end{abstract}

\pacs{42.50.Wk, 42.50.Pq, 07.10.Cm}

\maketitle
%======================================================
%                     Introduction
%======================================================
\section {Introduction}

The interplay between optical and mechanical degrees of freedom
via radiation pressure is at the heart of cavity optomechanics
\cite{review1,review3,review2,review4}. With the recent advances
in cooling techniques in optomechanical setups
\cite{cooling1,cooling2,cooling3,cooling4,cooling5,cooling6}, it
is becoming possible to access quantum ground states, and study
the interplay at the quantum level experimentally.  This may lead
to novel applications in quantum information based on
optomechanical coupling
\cite{Braunstein,Lukin,Wang,Mauro_Paternostro,genes_entanglement}.
Specifically, the system formed by a movable dielectric membrane
inside a optical cavity provides a basic configuration to explore
quantum phenomena in macroscopic objects ~\cite{membrane cooling,
membrane review, SiN film, membrane tunable, Meystre H,membrane
QND H}, such as non-classical mechanical states \cite{Meystre
squeezing H, membrane2 entang H} and cavity QED effects
~\cite{membrane with atom, membrane optomircomech}. Technically,
the system has the advantage that it enables a strong and tunable
optomechanical coupling, with the possibility of achieving
nonlinearity in the membrane's displacement~\cite{membrane
tunable}.

The theoretical analysis of a moving membrane system is mainly
based on the Hamiltonian ~\cite{Meystre H,membrane QND H, Meystre
squeezing H, membrane2 entang H,membrane tunable, membrane with
atom, membrane optomircomech}
\begin{eqnarray}
H
 & \approx & \hbar \omega(\hat{x}_{m}) a^{\dag} a
+ \hbar \Omega_{m} b^{\dag}b \nonumber \\
& \approx& \hbar \left( \omega_0 + \hat{x}_{m}
\left.\frac{\partial\omega(x)}{\partial x}\right|_{x_0} \right)
a^{\dag} a + \hbar \Omega_{m} b^{\dag}b,
\end{eqnarray}
where $a$ and $b$ are the annihilation operators for optical and
mechanical modes, respectively, $\hat{x}_{m} \propto (b+b^{\dag})$
is the displacement operator of dielectric motion, and $\Omega_m$
is the natural frequency of the membrane. The radiation pressure
coupling between the two degrees of freedom is contained in the
membrane-position-dependent field frequency $\omega(\hat{x}_{m})$,
which can be linearized around an equilibrium position $x_0$. If
the first derivative of field frequency vanishes, then the
coupling is dominantly quadratic in membrane displacements
\cite{membrane QND H,membrane tunable}.

We note that the essential assumption of the model (1) is that all
photons would stay in the same cavity mode throughout the
evolution because of the slow motion of the dielectric membrane.
This is an adiabatic single-mode approximation in which photon
scattering among different cavity modes are neglected. However, we
point out that the mode coupling induced by membrane's motion can
become significant in non-adiabatic regimes. This happens
especially when the oscillation frequency of the dielectric is
close to the frequency spacing between two cavity modes, in which
case transitions between the two modes can be resonantly enhanced.
Indeed, by exploiting such a kind of resonance, Dobrindt and
Kippenberg have recently indicated an optomechanical displacement
transducer with a high sensitivity \cite{kippenberg transducer}.
Hence a natural question of the moving-membrane system is how a
Hamiltonian model can be rigorously formulated, without employing
the adiabatic single-mode approximation. Such a Hamiltonian would
provide us with a basis of studying the quantum mechanics of
field-membrane systems. A better understanding of the
field-membrane interaction also opens up possibilities of new
schemes to manipulate quantum states of both the light field and
mechanical motion.

A key to address the field-membrane interaction is to treat both
the field and the moving membrane consistently as dynamical
variables. A similar problem for a moving perfect-mirror system
has been treated in Ref.~\cite{Law}. The case of moving-dielectric
system has yet been formulated, although the field Hamiltonian for
a dielectric or partially transparent mirror moving in a
prescribed trajectory has been discussed in the context of
dynamical Casimir effect \cite{Barton,salamone,Haro}. A major
conceptual difficulty of the problem is that the normal modes
associated with cavity field depend on the position of the
dielectric, which enters as a dynamical variable. These field
modes change with time as the dielectric membrane moves, which in
turn affect the radiation pressure on the membrane. A consistent
approach to the coupled field-membrane dynamics is hence essential
to tackle the problem. Recently, Biancofiore {\it et al.}
\cite{Vitali} have constructed a Hamiltonian based on a linearized
form of the radiation pressure coupling (i.e., to first order in
$x_m$) . While their Hamiltonian may also address non-adiabatic
photon scattering among cavity modes, it remains unclear how the
constructed Hamiltonian can be extended beyond the linear
coupling.

The main purpose of this paper is to provide a Hamiltonian
formulation that describes the optomechanical coupling without
using the single-mode adiabatic approximation and linear
approximation. Based on the interaction between macroscopic
dielectric and electromagnetic field, we first derive the
Lagrangian and Hamiltonian of the classical counterpart of the
system. The canonical quantization of the Hamiltonian and the
membrane-position dependent Fock states are then introduced. Our
Hamiltonian indicates how photons defined by such Fock states can
be coupled to various cavity modes through the motion of the
mirror. In the regime where adiabatic single-mode approximation
and linear approximation are applicable, our Hamiltonian can be
reduced to the usual form (1). Near the end of this paper we
indicate some physical consequences arising from the involvement
of the multiple cavity modes.

\section {The Classical Lagrangian and Equations of Motion}\label{sec: L & EOM}

We begin by considering a one-dimensional optical cavity of length
$l$ formed by two perfectly reflecting end mirrors. A movable
membrane of rigid uniform dielectric is placed inside the cavity.
The cavity field is specified by its vector potential ${\bf A} =
A(x,t){\bf e}_{z}$ $(0<x<l)$ under transverse gauge, with the
boundary conditions $A(x=0,t)=A(x=l,t)=0$. We assume
non-birefringent dielectric so that the two polarizations of the
field are decoupled, and hence it suffices to consider a linearly
polarized field. The dielectric is specified by its mass $m$, its
center-of-mass coordinate ${\bf q} = q(t){\bf e}_{x}$
$(d/2<q<l-d/2)$, and the dielectric constant
\begin{equation}
\epsilon(x,q) =
 \left\{ \begin{array}{ll}
            1+\chi,& q-d/2<x<q+d/2\\
            1,& \mbox{otherwise}
          \end{array}
 \right.
\label{eq:epsilon}
\end{equation}
where $d$ and $\chi$ are the width and susceptibility of the
dielectric, respectively. We have used the convention $\epsilon_0
= \mu_0 = 1$ (i.e., $c=1$), and assumed non-magnetic dielectric
$\mu = \mu_0$. We have also assumed a non-dispersive dielectric so
that $\epsilon$ does not depend on the field frequency.

\begin{figure}[t]
\centering
\includegraphics[width=7cm]{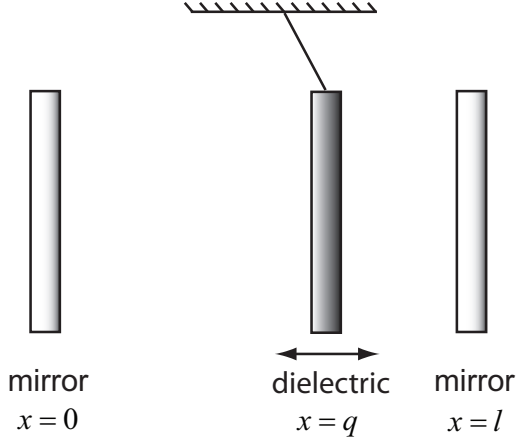}
\caption{The one-dimensional cavity with moving dielectric membrane at $x=q(t)$.}
\end{figure}

The motion of the dielectric affects the electromagnetic fields in
the cavity, which in turn modifies the radiation pressure on the
dielectric. To study the complete dynamics of the system, $q(t)$
must be included as a dynamical degree of freedom. The system is
specified by the Lagrangian
\begin{equation}
L = \frac{1}{2}m\dot{q}^2 - V(q) + \int_{0}^{l} dx {\cal L}_{F} ,
\label{eq:L start}
\end{equation}
where $V(q)$ is the external mechanical potential on the
dielectric, and ${\cal L}_{F}$ is the Lagrangian (linear) density
of the field after eliminating the electronic degrees of freedom
of the dielectric. To find ${\cal L}_{F}$, we go to an inertial
frame $S'$ in which the dielectric membrane is instantaneously at
rest. Assuming the acceleration of the membrane does not change
the macroscopic properties of the dielectric, the field Lagrangian
density in $S'$ is given by the familiar form: ${\cal L}_F' =
\frac{1}{2}\left( \epsilon {\bf E}'^2 - {\bf B}'^2
\right)$ %\label{eq:field L rest frame}
where ${\bf E}'=E'{\bf e}_{z}$ and ${\bf B}'=B'{\bf e}_{y}$ are
the electric and magnetic fields in $S'$, respectively. Now as the
motion of the dielectric in the laboratory frame $S$ is
perpendicular to both fields $\dot{{\bf q}}=\dot{q}(t){\bf
e}_{x}$, we can relate the fields between $S$ and $S'$ by the
Lorentz transformation: ${\bf E}' = \gamma \left({\bf E} +
\dot{{\bf q}}\times{\bf B}\right)$ and ${\bf B}' = \gamma
\left({\bf B} - \dot{{\bf q}}\times{\bf E}\right)$, with $\gamma =
\left(1-\dot{q}^{2}\right)^{-1/2}$. In terms of the vector
potential, ${\bf E} = -(\partial_{t}A) {\bf e}_{z}$ and $ {\bf B}
= -(\partial_{x}A){\bf e}_{y}$, the Lagrangian density ${\cal
L}_{F}$ in the space between the mirrors reads
\begin{equation}
{\cal L}_{F} = \frac{1}{2}\frac{1}{(1-\dot{q}^2)} \left[
\left(\epsilon-\dot{q}^2\right) \left(\frac{\partial A}{\partial
t}\right)^2 - \left(1- \epsilon \dot{q}^2\right)
\left(\frac{\partial A}{\partial x}\right)^2 + 2
\left(\epsilon-1\right)\dot{q} \left(\frac{\partial A}{\partial
t}\right)\left(\frac{\partial A}{\partial x}\right) \right].
\label{eq:field L barton}
\end{equation}
This agrees with the earlier work by Barton {\em et
al.}~\cite{Barton} and by Salamone~\cite{salamone} (in the case
$\mu=1$). We see that the Lagrangian density (\ref{eq:field L
barton}) appears to be more complicated compared with that in the
primed frame. This can be understood because in the laboratory
frame, a dielectric with polarization ${\bf P}$ moving at velocity
$\dot{{\bf q}}$ processes a magnetization ${\bf M} = - \dot{{\bf
q}} \times {\bf P}$, which must not be neglected in the regime of
non-adiabatic dielectric motion. After some calculations, it can
be shown that the Lagrangian density (\ref{eq:field L barton})
contains the interaction terms corresponding to ${\bf P}\cdot {\bf
E} + {\bf M}\cdot {\bf B}$.

In this paper we confine our study for non-relativistic motion
$\dot{q} \ll 1$, so that Eq.~(\ref{eq:field L barton}) is
approximated by (up to first order in $\dot{q}$),
\begin{equation}
{\cal L}_{F} =
\frac{1}{2} \left[\epsilon \left(\partial_{t} A\right)^2 -\left(\partial_{x} A\right)^2\right]
+ \dot{q} \left(\epsilon-1\right) \left(\partial_{t} A\right)\left(\partial_{x} A\right).
\label{eq:LFapprox}
\end{equation}
Together with Eq.~(\ref{eq:L start}), we obtain
\begin{equation}
L = \frac{1}{2}m\dot{q}^2 - V(q) + \int_{0}^{l} dx \ \{\frac{1}{2}
\left[\epsilon \left(\partial_{t} A\right)^2 -\left(\partial_{x}
A\right)^2\right] + \dot{q} \left(\epsilon-1\right)
\left(\partial_{t} A\right)\left(\partial_{x} A\right) \},
\label{eq:Lmodel}
\end{equation}
which is the non-relativistic Lagrangian of our membrane-field
model.

To justify this Lagrangian, we need to examine whether it
consistently generates the equations of motion for both the fields
and the membrane within the accuracy limited by the approximation
made in (\ref{eq:Lmodel}). First the Euler-Lagrange equation of
$A(x,t)$ derived from (\ref{eq:Lmodel}) is given by (to first
order in $\dot q$)
\begin{equation}
\epsilon \partial^{2}_t A - \partial^{2}_x A + 2 \dot{q} (\epsilon
-1) \left(\partial_{x}\partial_{t} A\right) + \ddot q (\epsilon
-1) \partial_x A   = 0, \label{eq:fieldEOM}
\end{equation}
where we have discarded terms involving $\dot q ^2$ for
consistency, and made use of the relation $(\partial_{t}+ \dot q
\partial_x) \epsilon =0$. From Eq. (\ref{eq:fieldEOM}), the effects of
membrane's motion on the field appear in the $\epsilon$ terms, the
third term with a velocity dependence, and the last term that is
proportional to the membrane's acceleration. If $\ddot q =0$, then
Eq.  (\ref{eq:fieldEOM}) is simply the wave equation obtained by
transforming the wave equation $\epsilon
\partial^{2}_{t'} A' - \partial^{2}_{x'} A'=0$ in $S'$ frame to $S$
frame up to first order in $\dot q$~\cite{Barton,Leonhardt}. The
acceleration dependent term therefore acts like a source term in
the wave equation. However, for an oscillating membrane with a
mechanical frequency $\Omega$ and field frequency $\omega$, the
ratio of the acceleration dependent term to the velocity dependent
term in Eq.  (\ref{eq:fieldEOM}) is of the order $\Omega/\omega$,
which is much smaller than one.

Next, the Euler-Lagrange equation of motion of the membrane based
on (\ref{eq:Lmodel}) is given by
\begin{equation}
m \ddot{q} = -\frac{\partial{V(q)}}{{\partial q}} + \frac{1}{2}
\frac{\chi}{1+\chi} \left[\left(\frac{\partial A}{\partial
x}\right)^2\right]^{q-d/2}_{q+d/2}. \label{eq:mirrorEOM}
\end{equation}
The second term on the right-hand side of Eq. (\ref{eq:mirrorEOM})
corresponds to a radiation pressure force from the field. Such a
force term is consistent with that obtained from the Lorentz force
density ${\bf f}'=(\partial_{t'} {\bf P}')\times {\bf B}'$
appearing in $S'$~\cite{Loudon}. This can be shown by using ${\bf
P}' = \chi {\bf E}'$ and the wave equation in dielectric rest
frame, then a straightforward transform on the force to the
laboratory frame in the non-relativistic limit would yield the
same radiation force expression in Eq. (\ref{eq:mirrorEOM}), apart
from a term that is about $\dot q/c$ times smaller. Note that our
Lagrangian (\ref{eq:Lmodel}) gives a wave equation that is
accurate up to $O(\dot{q})$, but the accuracy is lower by one
order of $\dot{q}$ for the membrane's equation of motion. This is
because of the partial derivative $\partial/\partial \dot{q}$ in
the membrane's Euler-Lagrange equation.

\section {The Hamiltonian and Quantization}

The Hamiltonian associated with $L$ is defined by
\begin{equation}
H\left(\Pi, A, p,q\right) \equiv p\dot{q} + \int_{0}^{l} dx \left[
{\Pi
\left(\partial_{t}A\right) - L\left(A,\partial_{t}A, q, \dot{q} \right)} \right],
\label{eq:H_def}
\end{equation}
where $p$ and $\Pi(x,t)$ are canonical momenta conjugate to $q$ and $A(x,t)$, respectively,
\begin{eqnarray}
p
&=& \frac{\partial L}{\partial \dot{q}}
= m \dot{q} + \int_{0}^{l} dx \left(\epsilon-1\right) \left(\partial_{t} A\right)\left(\partial_{x} A\right)
\label{eq:p_def}\\
\Pi
&=& \frac{\partial {\cal L}_{F}}{\partial (\partial_{t}A)}
=\epsilon\left(\partial_{t} A\right)
+ \dot{q} \left(\epsilon-1\right) \left(\partial_{x} A\right).
\end{eqnarray}
We see that the dielectric canonical momentum $p$ is not equal to
its kinetic momentum $m\dot{q}$ for non-zero fields. The explicit
expression of the Hamiltonian (\ref{eq:H_def}) now reads
\begin{equation}
H=\frac{1}{2m'} \left(p+\Lambda\right)^2 + V(q) + \frac{1}{2}
\int_{0}^{l} dx \left[\frac{\Pi^2}{\epsilon} + \left(\partial_{x}
A\right)^2\right], \label{eq:Hstart_m'}
\end{equation}
with $\Lambda$ given by
\begin{equation}
\Lambda = -\int_{0}^{l} dx \left(\frac{\epsilon-1}{\epsilon}\right) \Pi \left(\partial_{x} A\right),
\label{eq:Lambda unsymm}
\end{equation}
and $m'$ is identified as a `renormalized mass' defined by
\begin{equation}
m' = m\left[1-\frac{1}{m}\int_{0}^{l} dx \frac{\left(\epsilon-1\right)^2}{\epsilon}
\left(\frac{\partial A}{\partial x}\right)^2\right]^2.
\label{eq:m'_def_wrong}
\end{equation}
The form of Hamiltonian (\ref{eq:Hstart_m'}) is similar to the
minimal coupling Hamiltonian in electrodynamics with $\Lambda$
somehow playing the role of the vector potential in the kinetic
energy term.

At this point we would like to comment on the renormalized mass
$m'$ defined in Eq. (\ref{eq:m'_def_wrong}) \cite{jackel}. First,
it might look peculiar that the renormalized mass $m'$ depends
only on the magnetic field energy inside the membrane, but we
point out that this is an artifact due to the truncation of the
Lagrangian (\ref{eq:field L barton}) up to $\dot q$. If we retain
$\dot q^2$ terms in (\ref{eq:field L barton}), it can be shown
that $m'$ appears to depend on the electric field energy as well.
Second, in quantum theory, the vacuum field energy would make the
integral in Eq. (\ref{eq:m'_def_wrong}) divergent if all the field
frequencies are counted. In practice, however, a physical
dielectric membrane must become transparent (i.e. $\epsilon \to
1$) at high field frequencies, so there is only a finite range of
field frequencies contributing. It is useful to estimate the order
of magnitude if the field frequencies are counted up to
$\omega_{c}=10^{17}\mbox{ Hz}$ in the ultra-violet range. We then
find that for a $l=1\mbox{ cm}$ cavity, the vacuum contribution is
of the order $10^{-28}\mbox{ kg}$, many orders of magnitude
lighter than a pico-gram membrane in typical optomechanical setup.
If the cavity is filled with photon excitations in a single mode,
then a similar consideration shows that the photon number has to
be as high as $10^{15}$ for the mass correction to be comparable
to the mass of the dielectric. Hence the mass correction can be
safely neglected as long as we restrict our dielectric model to
optical field frequencies and a sufficiently massive membrane.
From now on, we will take $m'\approx m$, and the Hamiltonian reads
\begin{equation}
H=\frac{1}{2m} \left(p+\Lambda\right)^2 + V(q) + \frac{1}{2}
\int_{0}^{l} dx \left[\frac{\Pi^2}{\epsilon} + \left(\partial_{x}
A\right)^2\right]. \label{eq:Hstart_m}
\end{equation}

To quantize the system, we promote the dynamical variables $q,p,
A(x), \Pi(x)$ into operators by postulating the commutation
relations $[\hat{q}, \hat{A}(x)] = [\hat{q}, \hat{\Pi}(x)] =
[\hat{p}, \hat{A}(x)] = [\hat{p}, \hat{\Pi}(x)] = 0$,
$\left[\hat{q},\hat{p}\right] = i\hbar$, $[\hat{A}(x),
\hat{\Pi}(x')] = i\hbar \delta(x-x')$. The quantum Hamiltonian
takes the same expression as (\ref{eq:Hstart_m}), but with
$\Lambda$ defined in (\ref{eq:Lambda unsymm}) symmetrized as
\begin{equation}
\hat{\Lambda}(\hat{q}) = -\int_{0}^{l} dx \left(\frac{\epsilon-1}{2\epsilon}\right)
\left[\hat{\Pi} \left(\partial_{x} \hat{A}\right) + \left(\partial_{x} \hat{A}\right) \hat{\Pi}\right].
\label{eq:Lambda symm}
\end{equation}

\subsection {Instantaneous normal-mode projection}

\begin{figure}[t]
\centering
\includegraphics[width=5cm]{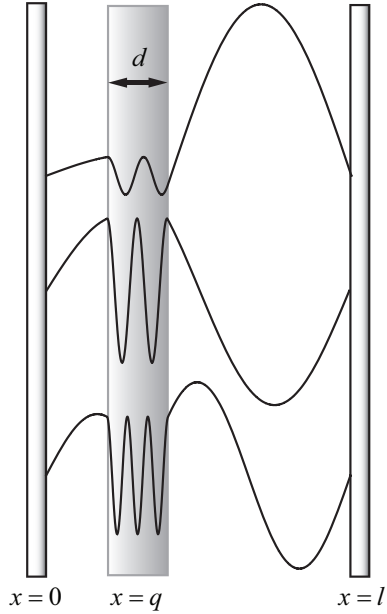}
\caption{A sketch of a few mode functions. Evidently the mode
functions depends on the position of the dielectric.}
\end{figure}

The field operators can be projected onto any set of complete
orthonormal modes. For instance, we may use the set of mode
functions $\{\varphi_{k}(x,q_0)\}$ defined by
\begin{equation}
\frac{\partial^2 \varphi_{k}(x,q_0)}{\partial x^2} +
\epsilon(x,q_0) \omega^{2}_{k}(q_0) \varphi_{k}(x,q_0) = 0
\label{eq:mode_ODE}
\end{equation}
with a vanishing boundary condition at $x=0$ and $x=l$. Here $q_0$
is a reference position (c-number), say, an equilibrium position
of the dielectric.  The orthonormality relation between these mode
functions is written as: $\int_{0}^{l} dx \epsilon(x,q_0)
\varphi_{k}(x,q_0) \varphi_{j}(x,q_0)= \delta_{kj}$. Note that we
have explicitly labelled the mode functions and frequencies as
$\varphi_{k}(x,q_0)$ and $\omega_{k}(q_0)$ to emphasize their
dependence on $q_0$, i.e., $\varphi_{k}(x,q_0)$ would be the
normal-mode of the field if the membrane had been {\em fixed} at
$x=q_0$.

By substituting $\hat{A}(x) = \sum_{k} \hat{Q}_{k}
\varphi_{k}(x,q_0)$ and $\hat{\Pi}(x) = \sum_{k} \hat{P}_{k}
\epsilon(x,q_0)\varphi_{k}(x,q_0)$, the Hamiltonian reads
\begin{equation}
H=\frac{1}{2m} \left[\hat{p}+\sum_{k,j} \frac{\xi_{kj}}{2}
\left( \hat{P}_{k} \hat{Q}_{j} + \hat{Q}_{j}\hat{P}_{k} \right)\right]^2
+ V(\hat{q}) + \frac{1}{2} \left(\sum_{k,j} \eta_{kj} \hat{P}_{k}\hat{P}_{j}
+ \sum_{k} \omega_{k}^2 \hat{Q}_{k}^2\right),
\label{eq:fixmodeH}
\end{equation}
where
\begin{eqnarray}
\hat{Q}_{k} &=& \int_{0}^{l} dx \epsilon(x,q_0)
\varphi_{k}(x,q_0)\hat{A}(x)
\label{eq:Q def}\\
\hat{P}_{k} &=& \int_{0}^{l} dx \varphi_{k}(x,q_0) \hat{\Pi}(x)
\label{eq:P def}\\
\eta_{kj}(\hat{q}) &=& \int_{0}^{l} dx \epsilon^{-1}(x,\hat{q}) \epsilon^{2}(x,q_0) \varphi_{k}(x,q_0)\varphi_{j}(x,q_0)\\
\xi_{kj}(\hat{q}) &=&
 -\int_{0}^{l} dx \left[\frac{\epsilon(x,\hat{q})-1}{\epsilon(x,\hat{q})}\right]  \epsilon(x,q_0)
 \varphi_{k}(x,q_0)\frac{\partial \varphi_{j}(x,q_0)}{\partial x}.
\end{eqnarray}
In this way the standard commutation relations: $[\hat{q},
\hat{Q}_{k}] = [\hat{q}, \hat{P}_{k}] = [\hat{p}, \hat{Q}_{k}] =
[\hat{p}, \hat{P}_{k}] = 0$, $[\hat{Q}_{j}, \hat{P}_{k}] = i\hbar
\delta_{kj}$ are preserved. Since $\varphi_{k}(x,q_0)$ is not an
instantaneous normal-mode of the cavity, the field part of the
Hamiltonian (\ref{eq:fixmodeH}) is not diagonalized.

To reveal the physical picture it is always desirable to cast the
field part of the Hamiltonian in a diagonal basis. This can be
achieved by performing a unitary transformation
$H'=T_{1}^{\dag}HT_{1}$ (where $T_1$ is defined in Appendix
\ref{details tran T1}):
\begin{equation}
H'
=\frac{1}{2m} \left(\hat{p}+\sum_{k,j} g_{kj}\hat{P}_k \hat{Q}_j\right)^2
+ V(\hat{q}) + \frac{1}{2}\sum_{k} \left[\hat{P}_{k}^{2} + \omega_{k}^{2}(\hat{q}) \hat{Q}_{k}^{2}\right],
\label{eq:instmodeH}
\end{equation}
with
\begin{equation}
g_{kj}(\hat{q}) = - \int_{0}^{l} dx \left\{ \epsilon(x,\hat{q})
\frac{\partial \varphi_j(x,\hat{q})}{\partial \hat{q}} +
[\epsilon(x,\hat{q})-1] \frac{\partial
\varphi_j(x,\hat{q})}{\partial x} \right \} \varphi_k(x,\hat{q})
\label{eq:gkj}
\end{equation}
which satisfies $g_{kj}=-g_{jk}$, so that $g_{kk}=0$. The field
part of $H'$ is now diagonalized with $\hat{q}$-dependent
frequencies $\omega_{k}(\hat{q})$. In particular, the transformed
field operators are given by
\begin{eqnarray}
T_{1}^\dag \hat{A}(x) T_{1} &=& \sum_{k} \left(T_{1}^\dag
\hat{Q}_{k}T_{1}\right) \varphi_{k}(x,q_0) = \sum_{k} \hat{Q}_{k}
\varphi_{k}(x,\hat{q})
\label{eq:T1 A}\\
T_{1}^\dag \hat{\Pi}(x) T_{1} &=& \sum_{k} \left(T_{1}^\dag
\hat{P}_{k}T_{1}\right) \epsilon(x,q_0)\varphi_{k}(x,q_0) =
\sum_{k} \hat{P}_{k} \epsilon(x,\hat{q})\varphi_{k}(x,\hat{q})
\label{eq:T1 Pi}
\end{eqnarray}
according to Appendix \ref{details tran T1}. $\varphi_{k}(x,\hat{q})$
is defined in Eq. (\ref{eq:mode_ODE}) but with
$q_0$ replaced by $\hat q$. Noting that $\hat q$
is the position operator of the membrane, $\varphi_{k}(x,\hat{q})$
corresponds to an {\em instantaneous normal-mode} of the cavity.
Therefore $\hat{Q}_{k}$ and
$\hat{P}_{k}$ become the expansion of {\em transformed}
$\hat{A}(x)$ and $\hat{\Pi}(x)$ using the instantaneous
normal-modes respectively:
\begin{eqnarray}
\hat{Q}_{k} &=& \int_{0}^{l} dx \left[T_{1}^{\dag}
\hat{A}(x)T_{1}\right] \epsilon(x,\hat{q}) \varphi_{k}(x,\hat{q})
\label{eq:new Q def}\\
\hat{P}_{k} &=& \int_{0}^{l} dx \left[T_{1}^{\dag}
\hat{\Pi}(x)T_{1}\right] \varphi_{k}(x,\hat{q}) \label{eq:new P def}.
\end{eqnarray}
It should be noted that $\hat{Q}_{k}$ and $\hat{P}_{k}$ are the
same operators defined in (\ref{eq:Q def}) and (\ref{eq:P def})
and hence they are independent of $\hat{q}$. Therefore, the
$\hat{q}$-dependence of $T_{1}(\hat{q})$, $\epsilon(x,\hat{q})$
and $\varphi_{k}(x,\hat{q})$ must have `cancelled out' each other
in (\ref{eq:new Q def}) and (\ref{eq:new P def}).

\subsection {Generalized Fock spaces}
\label{Fock space}

To represent the quantum state of the system, we introduce the
$\hat{q}$-dependent annihilation and creation operators for each
cavity field mode:
\begin{eqnarray}
a_{k}(\hat{q}) &=& \sqrt{\frac{1}{2\hbar \omega_{k}(\hat{q})}}
\left[\omega_{k}(\hat{q}) \hat{Q}_{k} + i \hat{P}_{k}\right]
\label{eq:a(q) def},\\
a^{\dag}_{k}(\hat{q}) &=& \sqrt{\frac{1}{2\hbar \omega_{k}
(\hat{q})}}\left[\omega_{k}(\hat{q}) \hat{Q}_{k} - i
\hat{P}_{k}\right] \label{eq:a^dag(q) def},
\end{eqnarray}
which satisfy the commutation relation $[a_{k}(\hat{q}),
a_{j}^{\dag}(\hat{q})] = \delta_{kj}$. Since $a_{k}(\hat{q})$
depends on $\hat{q}$, for each position of the dielectric we have
a set of Fock states associated with that position. These states
can be labelled as $|\{n_{k}\}, q\rangle$, where
$\{n_{k}\}=\{n_{1},n_{2},n_{3},\ldots\}$ denotes the occupation
number of each photon mode. $|\{n_{k}\}, q\rangle$ is a
simultaneous eigenstate of the photon-number operator
$a^{\dag}_{k}(\hat{q})a_{k}(\hat{q})$ and the position operator
$\hat{q}$ i.e. $ a^{\dag}_{k}(\hat{q})a_{k}(\hat{q})|\{n_{k}\},
q\rangle = n_{k} |\{n_{k}\}, q\rangle$ and $\hat{q} |\{n_{k}\},
q\rangle = q |\{n_{k}\}, q\rangle$. Such a set of eigenstates are
orthonormal and complete, so that any quantum state of the whole
system $|\Psi\rangle$ can be expanded in the basis of these
eigenstates i.e.
\begin{equation}
|\Psi\rangle = \sum_{\{n_{k}\}} \int_{d/2}^{l-d/2} C(\{n_{k}\},
q) |\{n_{k}\}, q\rangle dq,
\end{equation}
where $C(\{n_{k}\}, q)$ is the probability amplitude.

With the help of the $\hat{q}$-dependent annihilation and creation
operators, the Hamiltonian (\ref{eq:instmodeH}) becomes
\begin{equation}
H'=\frac{\left(\hat{p}+\Gamma\right)^2}{2m} + V(\hat{q}) +
\sum_{k} \hbar \omega_{k}(\hat{q}) \left(a^{\dag}_{k} a_{k}
+\frac{1}{2}\right), \label{eq:fock H vac}
\end{equation}
where we have used a shorthand $a_{k} = a_{k}(\hat{q})$ for
convenience, and
\begin{equation}
\Gamma(\hat{q}) = -\frac{i
\hbar}{2}\sum_{k,j}g_{k,j}(\hat{q})\sqrt{\frac{\omega_{k}}{\omega_{j}}}
\left(a_{k}a_{j} - a^{\dag}_{k}a^{\dag}_{j} + a^{\dag}_{j}a_{k} -
a^{\dag}_{k}a_{j}\right). \label{eq:Gamma}
\end{equation}
The vacuum field energy appearing in (\ref{eq:fock H vac}) leads
to the Casimir force on the dielectric (e.g. see~\cite{Casimir
dielectric} for calculations of a setup similar to our case). We
may replace the vacuum energy by the Casimir potential energy.
However, since the Casimir energy is feebly small compared with
$V(\hat{q})$ in typical optomechanical experiments, its effect
should be negligible.

\subsection {A unitary transformation}

The canonical momentum operator $\hat{p}$ in the Hamiltonian
(\ref{eq:fock H vac}) differs from the kinetic momentum $m\dot{q}$
due to $\Gamma(\hat{q})$. We may apply a unitary transformation to
the Hamiltonian to make the two momenta coincide. This procedure
is analogous to the case of atom-field interaction, where one
transforms the minimal-coupling Hamiltonian into the
electric-dipole form under electric-dipole approximation. The
transformation is defined with the unitary operator
\begin{equation}
T_{2}(\hat{q}) =\exp\left[-\frac{i}{\hbar}\int_{q_0}^{\hat{q}}
dq'\Gamma(q')\right]. \label{eq:T2 def}
\end{equation}
The transformed Hamiltonian $\tilde{H}=T_{2}^{\dag}H'T_{2}$ reads
(Appendix \ref{details tran T2})
\begin{equation}
\tilde{H} = \frac{\hat{p}^2}{2m} + u(\hat{q}) + \sum_{k} \hbar
\omega_{k}a^{\dag}_{k}a_{k} + \sum_{k,j} \hbar \left[
\xi_{kj}^{(+)}\left(a^{\dag}_{k}a_{j}+a^{\dag}_{j}a_{k}\right)
+\xi_{kj}^{(-)}\left(a^{\dag}_{k}a^{\dag}_{j}+a_{k}a_{j}\right)
\right], \label{eq:fock H tran}
\end{equation}
where
\begin{eqnarray}
u(\hat{q}) &=& V(\hat{q})
+ \frac{1}{2} \sum_{k} \hbar\left( \omega_{k}(\hat{q})+ 2\xi_{kk}^{(+)}(\hat{q})\right)\\
\xi_{kj}^{(\pm)}(\hat{q}) &=&
\frac{1}{4}\sqrt{\omega_{k}\omega_{j}} \left[ 2
\lambda_{kj}\frac{\omega_{k}}{\omega_{j}} + \sum_{l}
\lambda_{lk}\lambda_{lj}\frac{\omega^{2}_{l}}{\omega_{k}\omega_{j}}
\pm\left(\lambda_{kj}-\lambda_{jk}\right)
\right]\\
\lambda_{kj}(\hat{q}) &=& f_{kj} + \frac{1}{2!}\sum_{h}
f_{kh}f_{hj} + \frac{1}{3!}\sum_{h,l} f_{kh}f_{hl}f_{lj} + \ldots
\label{eq:lambda kj def}\\
f_{kj}(\hat{q}) &=& \int_{q_0}^{\hat{q}} dq' g_{kj}(q').
\end{eqnarray}
The Hamiltonian (\ref{eq:fock H tran}) is the main result of this
paper, which determines the coupling strengths $\xi_{kj}^{(\pm)}$
once the mode functions are known.  We stress that it is
applicable to general motion of the membrane, since no assumption
of the motion (except $\dot q \ll 1$) have been made. It should be
noted that the Hamiltonian (\ref{eq:fock H tran}) contains a
photon-number non-conserving part proportional to
$(a^{\dag}_{k}a^{\dag}_{j}+a_{k}a_{j} )$, which is responsible for
dynamical Casimir effect \cite{dodonov}.

We remark that the transformation has modified the mode function
associated with $a_{k}$, as the transformed cavity field operators
(related to that in (\ref{eq:Hstart_m}) by the combined unitary
transform defined via $T= T_{1}T_{2}$) read
\begin{eqnarray}
T^{\dag}\hat{A}(x)T &=& \sum_{k}
\sqrt{\frac{\hbar}{2\omega_{k}}}\left(a_{k}+a^{\dag}_{k}\right)
\tilde{\varphi}_{k}(x,\hat{q})
\label{eq:T A}\\
T^{\dag}\hat{\Pi}(x)T &=& -i\sum_{k}
\sqrt{\frac{\hbar\omega_{k}}{2}} \left(a_{k}-a^{\dag}_{k}\right)
\epsilon(x,\hat{q}) \tilde{\varphi}_{k}(x,\hat{q}) \label{eq:T
Pi}.
\end{eqnarray}
In other words, the mode functions have been changed to $
\tilde{\varphi}_{k}(x,\hat{q}) \equiv \varphi_{k}(x,\hat{q}) +
\sum_{j} \lambda_{jk}(\hat{q}) \varphi_{j}(x,\hat{q}) $ instead of
$\varphi_{k}(x,\hat{q})$. We show in Appendix \ref{details tran
T2} that $\{\tilde{\varphi}_{k}(x,\hat{q})\}$ indeed forms an
orthonormal complete set of mode functions.

\subsection {Linear approximation}

The Hamiltonian $\tilde{H}$ in (\ref{eq:fock H tran}) exhibits
nonlinear feature in the field-membrane coupling. In most
practical situations, the potential $V(\hat{q})$ bounds the
dielectric membrane about an equilibrium position $q_0$ such that
$\hat{x}_m = \hat{q} - q_0$ is small compared with the field
wavelengths concerned. Therefore we can perform the expansion
\begin{eqnarray}
\omega_{k}(\hat{q}) &\approx& \omega_{k0} + \hat{x}_{m}
\left(\left. \frac {\partial \omega_{k}} {\partial
q}\right|_{q=q_{0}}
\right)\\
a_{k}(\hat{q}) &\approx& a_{k0} + \frac{\hat{x}_{m}}{2\omega_{k0}}
\left(\left.\frac {\partial \omega_{k}} {\partial
q}\right|_{q=q_{0}}\right)
a^{\dag}_{k0}\\
\lambda_{kj} &\approx& \hat{x}_{m} g_{kj}^{(0)}
\end{eqnarray}
where $\omega_{k0}=\omega_{k}(q=q_{0})$ and $g_{kj}^{(0)}=
g_{kj}(q=q_{0})$ are the frequency and coupling constant
associated with the equilibrium position respectively, and
\begin{equation}
a_{k0} = \sqrt{\frac{1}{2\hbar \omega_{k0}}}\left(\omega_{k0}
\hat{Q}_{k} + i \hat{P}_{k}\right)
\end{equation}
is the annihilation operator linearized in $\omega_{k}(\hat{q})$.
The linearized $a_{k0}$ commutes with both $\hat{q}$ and
$\hat{p}$. The linearized $\tilde{H}$ reads
\begin{equation}
\tilde{H} = \frac{\hat{p}^2}{2m} + u(\hat{x}_m) + \sum_{k} \hbar \omega_{k0} a^{\dag}_{k0} a_{k0} + \hat{x}_{m} F_0,
\label{eq:multimode linearize H}
\end{equation}
where $F_0$ is the normal-ordered radiation pressure force
\begin{equation}
F_0 = \frac{\hbar}{2} \sum_{k,j} \left[ \left(\left.\frac{\partial
\omega_{k}}{\partial q}\right|_{q=q_0}\right) \delta_{kj} +
\omega_{k0} \sqrt{\frac{\omega_{k0}}{\omega_{j0}}}g^{(0)}_{kj}
\right] \left( a^{\dag}_{k0}a^{\dag}_{j0} + a_{k0}a_{j0} +
a^{\dag}_{k0}a_{j0} + a^{\dag}_{j0}a_{k0} \right).
\label{eq:multimode linearize F0}
\end{equation}
This agrees with the work in Ref. \cite{Vitali}, and the
corrections to the coupling term are of the order
$\hat{x}_{m}^{2}$.

\section {Some physical consequences of the Hamiltonian}

Current experimental and theoretical study on optomechanical
systems are mainly in the regime of single-mode and adiabatic
approximations. While our formulation does not require these
approximations, we show in this section how our Hamiltonian model
can be reduced to simpler forms when these approximations are
applicable. In addition, we indicate some possible modifications
due to the interaction with multiple cavity modes.

\subsection {Frequency shift in single-mode limit}

Single-mode approximation is applicable when the cavity field is
dominantly contributed by a single mode $k$, and photon
excitations in other modes are negligible. As $g_{kk}=0$, the
single-mode consideration immediately gives $\Gamma\approx 0$ from
(\ref{eq:Gamma}), and so the Hamiltonian (\ref{eq:fock H vac})
becomes
\begin{equation}
H'
\approx \frac{\hat{p}^2}{2m} + u(\hat{q})
+ \hbar \omega_{k}(\hat{q}) a^{\dag}_{k}(\hat{q})a_{k}(\hat{q}).
\label{eq:single mode rough}
\end{equation}
By linearizing in $\hat{x}_m = \hat{q} - q_0$ and employing
rotating-wave approximation (RWA), we have
\begin{equation}
H' \approx \frac{\hat{p}^2}{2m} + u(\hat{q}) + \hbar \omega_{k0}
a^{\dag}_{k0}a_{k0} + \hat{x}_{m} \left(\hbar \left.\frac{\partial
\omega_{k}}{\partial q}\right|_{q=q_0} \right)
 a^{\dag}_{k0}a_{k0},
\label{eq:single mode linearize rough}
\end{equation}
where $\omega_{k0}$ and $a_{k0}$ are the linearized frequency and
annihilation operator defined in the previous section.
Equivalently, (\ref{eq:single mode linearize rough}) can also be
obtained from (\ref{eq:multimode linearize H}) and
(\ref{eq:multimode linearize F0}) under single-mode approximation
together and neglecting the counter-rotating terms.

The above Hamiltonians (\ref{eq:single mode rough}) and
(\ref{eq:single mode linearize rough}) are what one would expect
when the dielectric motion is adiabatically slow. In obtaining
(\ref{eq:single mode rough}) we have simply `dropped out' the $j
\neq k$ terms in (\ref{eq:fock H vac}), hence scattering processes
between photon modes are neglected. However, it should be noted
that while only the field mode $k$ dominates, a photon in that
mode can make (virtual) transitions to other modes then back to
the $k$ mode. Such a process induces a shift in $\omega_{k}(q)$,
which is the leading order correction to (\ref{eq:single mode
rough}). To determine the correction, we work on our full
multi-mode Hamiltonian (\ref{eq:fock H tran}) and examine the
Heisenberg equation of motion for $a_{j}(\hat{q})$,
\begin{equation}
\dot{a}_{j}
= - i\omega_{j}a_{j}
+ \frac{1}{4}
\left(\dot{q}
\frac{1}{\omega_{j}}\frac{\partial \omega_{j}}{\partial q} a^{\dag}_{j}
+a^{\dag}_{j} \frac{1}{\omega_{j}}\frac{\partial \omega_{j}}{\partial q}\dot{q}
\right)
- i \sum_{l}
\left[
\left(\xi^{(+)}_{jl}+\xi^{(+)}_{lj}\right) a_{l}
+\left(\xi^{(-)}_{jl}+\xi^{(-)}_{lj}\right) a^{\dag}_{l}
\right].
\label{eq:EOM ak}
\end{equation}
We have used the equation of motion $\dot{q} = \hat{p}/m$ and the relation
\begin{equation}
\frac{\partial a_{j}}{\partial q}
= \frac{1}{2\omega_{j}}\frac{\partial \omega_{j}}{\partial q} a^{\dag}_{j}
\end{equation}
which follows from (\ref{eq:a(q) def}) and (\ref{eq:a^dag(q)
def}). The terms in (\ref{eq:EOM ak}) that contain $a^{\dag}_{j}$
and $a^{\dag}_{l}$ are fast-rotating, and can be neglected in the
spirit of rotating-wave approximation (RWA). Provided further that
the oscillation frequency of the membrane $\Omega$ is low compared
with the frequency difference $|\omega_{k}-\omega_{j}|$, we may
adiabatically eliminate $a_{j}$ $(j\neq k)$ in the equation of
motion of $a_{k}$:
\begin{equation}
\dot{a}_{k}
\approx - i\left[
\omega_{k} + 2\xi^{(+)}_{kk}
+ \sum_{\substack{k,j \\ j\neq k}}
\frac{\left(\xi^{(+)}_{k j}+\xi^{(+)}_{j k}\right)^2}
{\left(\omega_{k}+2\xi^{(+)}_{kk}\right)-\left(\omega_{j}+2\xi^{(+)}_{jj}\right)}
\right] a_{k}.
\end{equation}
Hence as a leading order non-adiabatic correction, the frequency
$\omega_{k}$ in (\ref{eq:single mode rough}) should be modified into
$\omega_{k}+\Delta_{k}$, where
\begin{equation}
\Delta_{k} = 2\xi^{(+)}_{kk}
+ \sum_{\substack{k,j \\ j\neq k}}
\frac{\left(\xi^{(+)}_{k j}+\xi^{(+)}_{j k}\right)^2}
{\left(\omega_{k}+2\xi^{(+)}_{kk}\right)-\left(\omega_{j}+2\xi^{(+)}_{jj}\right)}.
\label{eq:optical freq shift}
\end{equation}
We emphasize that this frequency shift is caused by the
interaction with other field modes, as is evident in the
expression of (\ref{eq:optical freq shift}). Hence the shift is
essentially a multi-mode effect, even though it is calculated
under the single-mode limit.

If the dielectric membrane is bounded by a harmonic potential
$u(\hat{x}_{m})\approx m \Omega^{2} \hat{x}_{m}^{2} /2$, with the
membrane equilibrium position $q_0$ at an extremum of
$\omega_{k}(\hat{q})$ (for example, at $q_0=l/2$), the linearized
single-mode Hamiltonian under RWA reads
\begin{eqnarray}
\tilde{H} &\approx& \frac{\hat{p}^2}{2m} + \frac{1}{2}m \Omega^{2}
\hat{x}_{m}^{2} + \hbar \omega_{k0} a^{\dag}_{k0}a_{k0} \nonumber \\
&&  + \hbar \hat{x}_{m}^{2} \left[ \frac{\omega_{k0}^{''}}{2} + \sum_{j}
g_{kj}^{(0)2} \frac
{\left(\omega_{k0}-\omega_{j0}\right)^{2}\left(\omega_{k0}+\omega_{j0}\right)}
{4\omega_{k0}\omega_{j0}} \right]
 a^{\dag}_{k0}a_{k0}
\label{eq:linearize shift H}
\end{eqnarray}
where $\omega_{k0}^{''} = \left(\partial^{2} \omega_{k}/\partial
q^2\right)_{q=q_0}$. The last term of (\ref{eq:linearize shift H})
comes from the Taylor expansion of $\Delta_{k}$, which has a
leading order of $\hat{x}_{m}^{2}$. It is intriguing to note that
the last term can be viewed in two ways: it can be regarded as a
shift in field frequency proportional to $\hat{x}_{m}^{2}$ (hence
proportional to phonon number under RWA \cite{membrane QND H}), as
well as a shift in the mechanical frequency $\Omega$ that is
proportional to the photon number.

Eq. (\ref {eq:linearize shift H}) shows that besides
$\omega_{k0}^{''}$, the frequency shift is also contributed by the
presence of other field modes. To compare the effect between these
two contributions, we have performed numerical calculations based
on parameters on the experimental setup of~\cite{membrane QND H}
where $l=0.06$ m, $n\equiv\sqrt{1+\chi}=2.2$, $d=50$ nm. With
$q_0=l/2$, a mode with $\omega_{k0}\approx 1.77\times 10^{15}$ Hz
(corresponding to $\lambda\approx 1064$ nm) is present with
$\omega_{k0}^{''}\approx -3.68 \times 10^{5}$~Hz~nm$^{-2}$. From
our calculation, modes with non-zero $g_{kj}^{(0)}$ contribute
about $0.22$~Hz~nm$^{-2}$ to the sum in~(\ref{eq:linearize shift
H}). While the contribution from each mode is feebly small, we
found that their contributions are approximately the same over a
wide spectral interval. Since these contributing modes have a
frequency spacing of about $3\times 10^{10}$ Hz, so if we include
modes of $10^{15} - 10^{16}$~Hz in our model, the correction would
sum to $\approx 0.7 \times 10^{5}$~Hz~nm$^{-2}$, which is only a
few times smaller than the contribution from $\omega_{k0}^{''}$.
However, we must point out that our result depends on the number
of modes included in the model. For this reason, the use of our
calculation is limited to an order-of-magnitude estimate. A more
detailed analysis should include the dispersion effect of the
dielectric, so as to address the spectral interval unambiguously.

\subsection {Resonant mode transitions}

Resonant mode transitions occur when the mechanical frequency of
the membrane $\Omega$ is comparable with the frequency spacing of
neighboring field modes. As an illustrative example, let us
consider the case where $q = q_{0} = l/2$ and a harmonic potential
$u(\hat{x}_{m})\approx m \Omega^{2} \hat{x}_{m}^{2} /2$. If the
index of refraction of the membrane is sufficiently high, the
eigen-frequencies of cavity modes distribute as doublets. Now
suppose the mechanical frequency $\Omega$ is close to the
frequency difference of two cavity modes in a doublet, say,
$k_{1}$ and $k_{2}$, then the two modes can be resonantly coupled.
Neglecting other non-resonant cavity modes, we approximate the
Hamiltonian (\ref{eq:multimode linearize H}) as
\begin{equation}
\tilde{H}
= \hbar \Omega \left(b^{\dag}b +\frac{1}{2}\right)
+ \hbar\left(\omega_{1} a^{\dag}_{1}a_{1} + \omega_{2} a^{\dag}_{2}a_{2}\right)
+ \hbar \eta \left(b+b^{\dag}\right) \left(a_{1}^{\dag} a_{2} + a_{2}^{\dag} a_{1}\right),
\label{eq:two mode H}
\end{equation}
where $\omega_{1}$ ($\omega_{2}$) and $a_{1}$ ($a_{2}$) are the
linearized frequency and annihilation operator of the $k_{1}$
($k_{2}$) mode, respectively, $b$ is the annihilation operator of
the dielectric motion, $\eta$ is the coupling frequency defined by
\begin{equation}
\eta = g^{(0)}_{12} \sqrt{\frac{\hbar}{8m\Omega}}
\frac
{\left(\omega_{1}^2 - \omega_{2}^2\right)}
{\sqrt{\omega_{1} \omega_{2}}}.
\end{equation}
Note that due to symmetry of the system,
\begin{equation}
\left.\frac{\partial \omega_{k_{1}}(q)}{\partial
q}\right|_{q=q_{0}} = \left.\frac{\partial
\omega_{k_{2}}(q)}{\partial q}\right|_{q=q_{0}} = 0,
\end{equation}
hence the usual radiation pressure term (\ref{eq:single mode
linearize rough}) is absent, and the field-membrane coupling to
first order in $\hat{x}_{m}$ describes the scattering between the
two field modes (as in \cite{kippenberg transducer}). In
particular, the case of $\Omega=\omega_2-\omega_1$ (assuming
$\omega_2
> \omega_1$) corresponds to a resonance at which the mode
coupling can be resonantly enhanced. If a rotating-wave
approximation is made, the interaction terms in Eq. (\ref{eq:two mode H})
would take the same form as that appears in parametric down conversion in
nonlinear optics.

As a remark we note that the field frequency spacing can be
reduced by increasing the cavity length or tuning the refractive
index and thickness of the membrane. Recent study \cite{membrane
tunable} also shows that avoided crossings of transverse field
modes (due to broken symmetry of the cavity along its lateral
dimensions) can provide a frequency spacing of the order $\sim1$
MHz, which is comparable to the mechanical frequency achievable in
current optomechanical experiments~\cite{58MHz}. Therefore it
would be possible for the membrane frequency to match the field
frequency spacing.

\section {Conclusion}
To conclude, we have presented a non-relativistic Lagrangian and a
Hamiltonian for a one-dimensional coupled membrane-field system in
the $\dot q \ll 1$ regime. The classical equations of motion of
both the field and membrane are obtained within the accuracy
limited by the approximation made in the Lagrangian. Our
Hamiltonian (\ref{eq:fock H tran}) should capture  optomechanical
processes not described by the single-mode adiabatic approximation
and linear approximation. For example, we have indicated that the
presence of multiple cavity-modes can modify the single-mode
Hamiltonian [see Eq. (\ref{eq:linearize shift H})], and possibly give rise to parametric
down conversion type interaction if the membrane frequency matches
the frequency spacing of cavity modes. With the explicit form of
interaction strengths between the membrane and various cavity
modes in (\ref{eq:fock H tran}), one can further study quantum
dynamics resulted from optomechanical coupling in non-adiabatic
regimes.

There are interesting subtle features that emerge in developing
the Hamiltonian model. First, the velocity-dependent coupling in
the Lagrangian (\ref{eq:LFapprox}), which is necessary to recover
the leading radiation pressure force term in (\ref{eq:mirrorEOM}),
causes the membrane to have a canonical momentum different from
its kinetic momentum. The unitary transformation $T_2$ can
eliminate the difference, but it turns out that $T_2$ also changes
the field operators accordingly. In this paper we have shown that
the transformation $T_2$ on the fields can be interpreted as a
modification of mode functions. Another subtle feature is the use
of instantaneous mode functions in diagonalizing the field part of
the Hamiltonian (\ref{eq:instmodeH}). The Fock states associated
with instantaneous modes are defined by the $\hat{q}$-dependent
photon creation/annihilation operators. It may not be convenient
to perform calculations directly based on such creation and
annihilation operators, but these operators can always be expanded
around the equilibrium position of the membrane in order to obtain
the relevant interaction terms. Using our approach, such an
expansion can be carried out to the first order of $\hat{x}_{m}$
(as in Sec. IIID), or to higher orders as desired. Finally, we
should point out that since our model is based on the
non-dispersive approximation of the dielectric, a further
investigation should incorporate the absorptive and dispersive
properties. This requires an extension of the current theory of
quantized field in dispersive media \cite{hut} to a moving media.

\vspace{5mm} \emph{Acknowledgments}--This work is partially
supported by a grant from the Research Grants Council of Hong
Kong, Special Administrative Region of China (Project
No.~CUHK401810).

\appendix
\section {The $T_{1}(\hat{q})$ transformation}
\label{details tran T1}
The unitary operator $T_{1}(\hat{q})$ is defined as
\begin{equation}
T_{1}(\hat{q}) =
\exp\left[-\frac{i}{\hbar}\sum_{k,j}\int_{q_0}^{\hat{q}} dq'
\zeta_{kj}(q') \frac{\left( \hat{P}_{k} \hat{Q}_{j} +
\hat{Q}_{j}\hat{P}_{k} \right)}{2}\right], \label{eq:expT}
\end{equation}
with
\begin{equation}
\zeta_{kj}(q') = \int_{0}^{l} dx \epsilon(x,q') \frac{\partial \varphi_{j}(x,q')}{\partial q'} \varphi_{k}(x,q').
\end{equation}
Here $\varphi_{k}(x,q')$ are mode functions defined in Eq. (\ref{eq:mode_ODE}) but with
$q_0$ replaced by $q'$. The integrand $\zeta_{kj}(q')$ in (\ref{eq:expT})
 is a c-number except at the upper limit
$\hat{q}$. In other words, the integral in the exponential of
(\ref{eq:expT}) can be viewed as an anti-derivative of
$\zeta_{kj}(q')$, evaluated at the two end-points. It follows that
\begin{equation}
p' = T_{1}^{\dag}\hat{p}T_{1}
= \hat{p} -i\hbar T_{1}^{\dag} \frac{\partial T_{1}}{\partial \hat{q}}
= \hat{p} - \sum_{k,j}  \frac{\zeta_{kj}}{2} \left( \hat{P}_{k} \hat{Q}_{j} + \hat{Q}_{j} \hat{P}_{k} \right).
\label{eq:p tilde}
\end{equation}
To consider the transformation on $\hat{Q}_{k}$ and $\hat{P}_{k}$, we decompose $T_{1}$ into a (continuous) product of infinitesimal transform
\begin{equation}
T_{1} = \prod_{q'=q_0}^{\hat{q}} \left[I - \frac{i}{\hbar} dq'
\sum_{k,j} \zeta_{kj}(q')\frac{\left(\hat{P}_{k} \hat{Q}_{j} +
\hat{Q}_{j}\hat{P}_{k} \right)}{2} \right] \equiv
\prod_{q'=q_0}^{\hat{q}} \left(I - \frac{i}{\hbar} dq'
K(q')\right) \label{eq:prodT}
\end{equation}
where the product is `$q'$-ordered', i.e. the leftmost term is
associated with $q'=q_0$, the rightmost with $q' = \hat{q}$. One
can check that (\ref{eq:expT}) and (\ref{eq:prodT}) are equivalent
by expanding the product of (\ref{eq:prodT}) into sums of
integrals. Each infinitesimal transform acts on $\hat{Q}_{k}$ and
$\hat{P}_{k}$ as
\begin{eqnarray}
\left(I + \frac{i}{\hbar} dq' K(q')\right) \hat{Q}_{k} \left(I - \frac{i}{\hbar} dq' K(q')\right)
&=& \hat{Q}_{k} + dq' \sum_{j} \zeta_{kj}(q') \hat{Q}_{j}\\
\left(I + \frac{i}{\hbar} dq' K(q')\right) \hat{P}_{k} \left(I - \frac{i}{\hbar} dq' K(q')\right)
&=& \hat{P}_{k} - dq' \sum_{j} \zeta_{jk}(q') \hat{P}_{j}
\end{eqnarray}
with correction only in second order infinitesimals. Hence the
infinitesimal transform modifies $\hat{A}(x)$ and $\hat{\Pi}(x)$ by
\begin{eqnarray}
\nonumber &&\left(I + \frac{i}{\hbar} dq' K(q')\right) \left(\sum_{k} \hat{Q}_{k} \varphi_{k}(x,q')\right) \left(I - \frac{i}{\hbar} dq' K(q')\right)\\
\nonumber &=& \sum_{k} \hat{Q}_{k}\varphi_{k}(x,q') + dq' \sum_{k,j} \hat{Q}_{j} \zeta_{kj}(q') \varphi_{k}(x,q')\\
\nonumber &=& \sum_{k} \hat{Q}_{k} \left[\varphi_{k}(x,q') + dq' \frac{\partial \varphi_{k}(x,q')}{\partial q'}  \right]\\
&=& \sum_{k} \hat{Q}_{k} \varphi_{k}(x,q'+dq'),
\end{eqnarray}
and
\begin{eqnarray}
\nonumber &&\left(I + \frac{i}{\hbar} dq' K(q')\right)
\left(\sum_{k}  \hat{P}_{k} \epsilon(x,q')\varphi_{k}(x,q')\right) \left(I - \frac{i}{\hbar} dq' K(q')\right)\\
\nonumber &=& \sum_{k} \hat{P}_{k} \epsilon(x,q')\varphi_{k}(x,q') - dq' \sum_{k,j} \hat{P}_{j} \epsilon(x,q') \zeta_{jk}(q') \varphi_{k}(x,q')\\
\nonumber &=& \sum_{k} \hat{P}_{k}
\left[\epsilon(x,q')\varphi_{k}(x,q') + dq' \frac{\partial}{\partial q'} \left(\epsilon(x,q')\varphi_{j}(x,q')\right)\right] \\
&=& \sum_{k} \hat{P}_{k} \epsilon(x,q'+dq')\varphi_{k}(x,q'+dq').
\end{eqnarray}
We have used the completeness relation
\begin{equation}
\sum_{k} \epsilon(x,q') \varphi_{k}(x,q')\varphi_{k}(x',q') = \delta(x,x'),
\end{equation}
and integration by parts
\begin{eqnarray}
\int_{0}^{l} dx' \epsilon(x',q') \frac{\partial \varphi_{j}(x',q')}{\partial q'} \varphi_{k}(x',q')
= - \int_{0}^{l} dx' \frac{\partial}{\partial q'} \left[\epsilon(x',q') \varphi_{k}(x',q') \right] \varphi_{j}(x',q').
\end{eqnarray}
It follows that combining all the infinitesimal transform readily
gives (\ref{eq:T1 A}) and (\ref{eq:T1 Pi}), which, together with
(\ref{eq:p tilde}), lead to (\ref{eq:instmodeH}). Note that we
need not symmetrize $\hat{P}_{k}\hat{Q}_{j}$ in (\ref{eq:instmodeH}) because
$g_{kj}=-g_{jk}$, and hence $g_{kk}=0$. This assertion can be
proven by performing integration by parts on (\ref{eq:gkj}), and
noting that the dielectric constant $\epsilon$ is fixed in its
rest frame, so that $\partial_{t'} \epsilon = \gamma\left(\partial_{t} +
\dot{q}\partial_{x}\right)\epsilon = \gamma\dot{q} \left(\partial_{q} +
\partial_{x}\right)\epsilon = 0$,
leading to $\partial_{q}\epsilon = -\partial_{x}\epsilon$.

\section {The $T_{2}(\hat{q})$ transformation}
\label{details tran T2}
With the unitary transform
$T_{2}(\hat{q})$ defined in (\ref{eq:T2 def}), the canonical
momentum operator $\hat{p}$ transforms as
\begin{equation}
T^{\dag}_{2}\hat{p}T_{2} = \hat{p} - \sum_{k,j} g_{kj}(\hat{q})\hat{P}_k \hat{Q}_j = \hat{p} - \Gamma
\end{equation}
so that $p$ becomes the kinetic momentum $m\dot{q}$ in the new
basis. The transform field variables $\tilde{Q}_{k} =
T^{\dag}_{2}\hat{Q}_{k}T_{2}$ and $\tilde{P}_{k} =
T^{\dag}_{2}\hat{P}_{k}T_{2}$ reads
\begin{eqnarray}
\tilde{Q}_{k} &=& \hat{Q}_{k} + \sum_{j} \lambda_{kj} \hat{Q}_{j}
\label{eq:tilde Q}\\
\tilde{P}_{k} &=& \hat{P}_{k} + \sum_{j} \lambda_{kj} \hat{P}_{j}
\label{eq:tilde P}
\end{eqnarray}
with $\lambda_{kj}$ defined in (\ref{eq:lambda kj def}). Hence the
fields in the new frame (i.e. $T^{\dag}A(x)T$ and $T^{\dag}\Pi(x)T$)
can be expanded in terms of $\hat{Q}_{k}$ and $\hat{P}_{k}$ by (\ref{eq:T A})
and (\ref{eq:T Pi}) respectively.

Next, we show that $\tilde{\varphi}(x,\hat{q})$ is orthonormal and
complete. Using the fact
$[\tilde{Q}_{j},\tilde{P}_{k}]=[\hat{Q}_{j},\hat{P}_{k}]=i\hbar \delta_{kj}$
for unitary transformation, together with (\ref{eq:tilde Q}) and
(\ref{eq:tilde P}), we have the identity
\begin{equation}
\lambda_{kj} + \lambda_{jk} + \sum_{l} \lambda_{kl}\lambda_{jl} = 0.
\end{equation}
Furthermore, by examining the form of $\lambda_{kj}$ with $f_{kj}
= -f_{jk}$, it can be shown that
\begin{equation}
\sum_{l} \lambda_{kl}\lambda_{jl} = \sum_{l} \lambda_{lk}\lambda_{lj}.
\end{equation}
These two properties of $\lambda_{jk}$ readily lead to
\begin{eqnarray}
\int_{0}^{l} dx \epsilon(x,\hat{q}) \tilde{\varphi}_{k}(x,\hat{q}) \tilde{\varphi}_{j}(x,\hat{q})
&=& \int_{0}^{l} dx \epsilon(x,\hat{q}) \varphi_{k}(x,\hat{q}) \varphi_{j}(x,\hat{q})
= \delta_{kj},\\
\sum_{k} \epsilon(x,\hat{q}) \tilde{\varphi}_{k}(x,\hat{q})\tilde{\varphi}_{k}(x',\hat{q})
&=& \sum_{k} \epsilon(x,\hat{q}) \varphi_{k}(x,\hat{q})\varphi_{k}(x',\hat{q})
= \delta(x,x'),
\end{eqnarray}
hence $\{\tilde{\varphi}_{k}(x,\hat{q})\}$ indeed forms an orthonormal
complete set of mode functions.

\end{document}